\documentclass{sig-alternate}

\usepackage{amsmath}
\usepackage{mathtools}		
\usepackage{fixltx2e}		
\usepackage{graphicx}		
\usepackage{subcaption}
\usepackage{bm} 			
\usepackage{algpseudocode}	
\usepackage{algorithm}		
\usepackage{tikz}
\usetikzlibrary{positioning,shapes.geometric}

\def\sharedaffiliation{%
\end{tabular}
\begin{tabular}{c}}

\usepackage[]{todonotes}

\begin{document}
%
\conferenceinfo{SC15, 27th International Conference for High Performance Computing, Networking, Storage and Analysis}{Austin, TX, USA}

\title{Task-Based Algorithm for Matrix Multiplication: A Step Towards Block-Sparse Tensor Computing}
%
%
%
%
%

\numberofauthors{2} 
%
\author{
%
%
  \alignauthor Justus A. Calvin\\
  \email{justusc@vt.edu}
  \alignauthor Edward F. Valeev\\     
  \email{efv@vt.edu}
  \sharedaffiliation
  \affaddr{Department of Chemistry}  \\
  \affaddr{Virginia Tech}\\
  \affaddr{Blacksburg, VA}\\
}

\maketitle

\begin{abstract}
%
%

%
Distributed-memory matrix multiplication (MM) is a key element of algorithms in many domains (machine learning, quantum physics). Conventional algorithms for dense MM rely on regular/uniform data decomposition to ensure load balance. These traits conflict with the irregular structure (block-sparse or rank-sparse within blocks) that is increasingly relevant for fast methods in quantum physics. To deal with such irregular data we present a new MM algorithm based on Scalable Universal Matrix Multiplication Algorithm (SUMMA). The novel features are: (1) multiple-issue scheduling of SUMMA iterations, and (2) fine-grained task-based formulation. The latter eliminates the need for explicit internodal synchronization; with multiple-iteration scheduling this allows load imbalance due to nonuniform matrix structure. For square MM with uniform and nonuniform block sizes (the latter simulates matrices with general irregular structure) we found excellent performance in weak and strong-scaling regimes, on commodity and high-end hardware.

\end{abstract}



\keywords{matrix multiplication, tensor contraction, task parallelism, SUMMA, latency tolerance, distributed computation}

\section{Introduction}

High performance parallel algorithms for matrix multiplication (MM) --- the most important special case of the general tensor contraction, and often its building block --- have been studied for decades \cite{Cannon:1969vd,Fox:1987ve,Berntsen:1989bx,Lin:1992dn,HussLederman:1993gp,Choi:1994jw,VanDeGeijn:1997uf,Gunnels:1998gg}.
MM nevertheless continues to draw attention of researchers \cite{Solomonik:2011bra,Tan:2013fl,Ballard:2012ir,Demmel:2013hy,Lai:2013el,Ozog:2013fu,Solomonik:2014ko,Rajbhandari:2014ue,Rajbhandari:2014bx,Rubensson:2015ur} due to the continuing evolution of the computer hardware as well as the prominent role of matrix and tensor computation in a variety of scientific domains, such as physics of classical and quantum fields (most notably, electronic structure \cite{Baumgartner:2002:HLA,Hirata:2003is,Epifanovsky:2013dl,Solomonik:2014ko}) as well as data analysis and model building in machine learning \cite{Hutchinson:2012bt}, chemometrics \cite{Smilde:2004wf}, neuroscience \cite{Morup:2008je}, and many more.
A frontier challenge posed by the scientific domain needs is the increasing importance of sparse and non-standard tensorial data representations (dense and sparse tensors, multiresolution spectral element trees, H matrices, matrix product states, tensor networks, {\em etc.}).
Another major challenge is the increasing complexity of communication hierarchy and the continuing increase 
of the communication cost, relative to that of computation.
This spurs the search for algorithms that minimize/avoid communication and/or hide its cost.
Here we explore improvements of a standard dense matrix-multiplication algorithm
that can hide communication costs and tolerate network latencies, data inhomogeneity, and sparsity, which serves as a platform for the development of algorithms on irregular tensorial data structures.

To demonstrate the tension between standard dense MM algorithms and the needs of emerging scientific domains, consider
a concrete example of chemistry and materials science. The matrices in such context represent quantum states (of electrons) and operators represented in some basis. Efficient application of operators to states --- represented by matrix multiplication --- demands taking advantage of the matrix structure arising from the physical properties of the entities these matrices represent. Such structure could be (a) block-sparsity due to the distance decay of the operator kernel and the localized nature of basis functions \cite{Hollman:2015ca}, (b) symmetries under geometric and other transformations \cite{Rajbhandari:2014bx,Hirata:2003is}, or (c) block-rank-sparsity due to near-sightedness of physical interactions \cite{Neese:2009dz,Neese:2009kd,Pavosevic:2014ed} ({\em i.e.} blocks of the matrix/tensor are dense but have low-rank representations; in the applied math community related matrix structures are known as H matrices).
Notably, the matrix structure is affiliated with the problem-specific blocking of matrix dimensions that arises due to domain-specific needs and often cannot be chosen arbitrarily.
In other words, the matrices that we encounter are ``sparse'' in a general sense, which encompasses element, block, and block-level-rank sparsity; but in a practical sense the matrices are not sparse enough to be a good match for sparse MM algorithms.
Nonuniform blocking and inhomogeneity of data due to the matrix structure conflict with the {\em uniform} data distribution of
standard distributed-memory dense MM algorithms --- including Cannon's \cite{Cannon:1969vd}, SUMMA \cite{VanDeGeijn:1997uf}, and others \cite{Fox:1987ve,Choi:1994jw,Buluc:2008vn,Solomonik:2014ko,Rajbhandari:2014bx}.

To design algorithms for multiplication of matrices with an irregular structure, we propose a reformulation of standard, dense MM algorithms that allows them to be tolerant of data inhomogeneity by expressing them in terms of tasks.
By using the standard algorithm as the framework our approach will be potentially optimal for dense uniformly-blocked matrices, and yet capable of handling irregularity in the matrix structure.
Task-based/dataflow programming models are a natural choice for implementation of algorithms with irregular data and computation patterns; such models have been used successfully for dense matrix algebra applications \cite{Husbands:2007de,pdsec}.
Besides handling matrices with structure, a task based approach will provide
additional benefits: (a) inter-node communication costs can be partially or fully hidden by overlapping computation and communication, (b) performance should be less sensitive to topology and tolerant of latency and CPU clock variations, (c) fine-grained task-based parallelism is a proven means to attain high intra-node performance by leveraging massively multicore platforms and hiding the costs of memory hierarchy ({\em e.g.}\ Intel TBB, Cilk), (d) lack of global synchronization allows the overlap multiple high-level stages of computation ({\em e.g.}\ multiple matrix multiplications contributing to the same expression).

This paper describes our approach, its implementation, and performance for matrix multiplication of square matrices with uniform and nonuniform blocking. Our implementation is available as part of the open-source project TiledArray (a general-purpose tensor library) \cite{TiledArray}.

\section{Distributed-Memory Matrix Multiplication}

This section contains an overview of the relevant literature on dense matrix multiplication algorithms for distributed memory computers.
Our discussion does not consider {\em fast} algorithms \cite{Cohn:2003em}, e.g. Strassen \cite{Strassen:1969im}, that have computational complexity lower than $\Theta(N^{3})$ of the na\"ive algorithm, where $N$ is the matrix size.
Thus for our purposes, all algorithms are computation-optimal and only differ in the algorithm construction, communication patterns, and memory usage per node.

An early parallel matrix-multiplication algorithm that has gained widespread use is Cannon's algorithm.
In this method, the input and output matrices are embedded onto a two-dimensional (2D) mesh of $P$ processes (nodes), where the input matrices are moved via a series of row and column rotations in systolic loops \cite{Cannon:1969vd}.
Cannon's algorithm asymptotically meets the lower bound on communication, $\Omega(N^{2}/\sqrt{P})$ \cite{Irony:2004kz}, for 2D algorithms that require ${\cal O}(N^{2}/P)$ memory per process.
The broadcast-multiply-roll algorithm of Fox {\em et al.}, which is also asymptotically optimal, introduced the use of a (pipelined) broadcast for movement of one of the two input matrices \cite{Fox:1987ve,Fox:1988ws}.
Unfortunately neither algorithm is universally applicable \cite{VanDeGeijn:1997uf,Schatz:2012wy}.
Some of the limitations were removed in the Transpose algorithm of Lin and Snyder \cite{Lin:1992dn}, and further by Choi and coworkers in Parallel Universal Matrix Multiplication Algorithm (PUMMA) \cite{Choi:1994jw} and by Huss-Lederman {\em et al.} in the BiMMeR algorithm \cite{HussLederman:1993gp}.
Perhaps the simplest and easiest to generalize is a group of 2D algorithms utilizing row/column broadcasts followed by rank-$k$ updates.
These algorithms were developed independently by several groups \cite{Agarwal:1994hj,Choi:1996dv,VanDeGeijn:1997uf}, most notably by van de Geijn and Watts who dubbed the approach Scalable Universal Matrix Multiplication Algorithm (SUMMA) \cite{VanDeGeijn:1997uf}.
The simplicity and flexibility of SUMMA made it very successful ({\em e.g.}\ it is the standard implementation of GEMM in ScaLAPACK), and has motivated several variants \cite{Choi:1997cp,Gunnels:1998gg,Buluc:2008vn}.
In addition, the core ideas are also widely used in linear algebra algorithms \cite{Poulson:2010wx,Solomonik:2011bra,Solomonik:2014ko,Rajbhandari:2014bx}.
Since our work is based on SUMMA, it is described in more detail in Section~\ref{ssec:SUMMA}.

All 2D algorithms described so far are optimal with respect to memory use.
However, it is possible to reduce the communication cost further by using a three-dimensional (3D) mesh of processes at the cost of additional memory usage \cite{Irony:2004kz,Ballard:2012dw}.
In 3D algorithms \cite{Berntsen:1989bx,Agarwal:1995jl}, each matrix is replicated across one dimension of a $\sqrt[3]{P} \times \sqrt[3]{P} \times \sqrt[3]{P}$ mesh.
This data distribution reduces the communication cost by a factor of $P^{1/6}$ \cite{Ballard:2012dw}, but requires ${\cal O}(N^{2}/P^{2/3})$ data per node, a factor of $P^{1/3}$ increase over the 2D case.
The gap between the 2D and 3D algorithms is bridged by the so-called 2.5D algorithms \cite{Solomonik:2011bra} that are communication-optimal with and without memory constraints \cite{Ballard:2012dw}.

Although 2D MM algorithms are not universally commu-nication-optimal, we view them as key for two reasons:
(a) they are the only choice under tight memory constraints, which is often the case in data intensive applications, 
and (b) they are a building block for higher-dimensional MM algorithms \cite{Solomonik:2011bra}.
Thus in our work we focus on SUMMA as the most popular 2D algorithm.

\subsection{\label{ssec:SUMMA} 2D SUMMA and Its Variants}
SUMMA implements matrix multiplication {\bf C} = {\bf A}{\bf B} as a series of rank-$k$ updates.
The input and output matrices are embedded on a rectangular process mesh in an element-cyclic or block-cyclic manner to ensure approximate load balance.
In each iteration of the algorithm a column/row panel of {\bf A}/{\bf B} is broadcast along rows/columns of the process grid, respectively; matrix {\bf C} remains stationary throughout the procedure (variants of SUMMA in which {\bf A} or {\bf B} are stationary are also possible; transposed multiplies, e.g. {\bf C} = {\bf A}{\bf B}$^{\dagger}$, are also relatively simple to handle \cite{VanDeGeijn:1997uf,Schatz:2012wy}).
Each pair of broadcasts is followed by a rank-$k$ update, $C_{ij} \gets A_{ik} B_{kj} + C_{ij}$, (Einstein summation convention is used throughout).
Original SUMMA papers by van de Geijn and Watts \cite{VanDeGeijn:1997uf} and by Agarwal {\em et al.} \cite{Agarwal:1994hj} considered versions of the algorithm that overlapped computation and communication by pipelining and preemptive broadcasts, respectively, and other broadcast variants have been considered \cite{Schatz:2012wy}, including topology-specific broadcasts \cite{Solomonik:2011jm}.
For simplicity, we present a SUMMA version with preemptive broadcasts in Figure~\ref{alg:basesumma}.

\begin{algorithm}[h]
  \caption{SUMMA with non-blocking broadcast}
  \begin{algorithmic}[1]
    \State \Call{Broadcast}{$A(*,0), 0, row\_group$}
    \State \Call{Broadcast}{$B(0,*), 0, col\_group$}
    \For{$k = 0, \dots, K-1$}
      \If {$k+1 < K$}
        \State $row\_root \gets (k+1) \bmod cols$
        \State $col\_root \gets (k+1) \bmod rows$
        \State \Call{Broadcast}{$A(*,k+1), row\_root, row\_group$}
        \State \Call{Broadcast}{$B(k+1,*), col\_root, col\_group$}
      \EndIf
      \State \Call{Wait}{$A(*,k)$}
      \State \Call{Wait}{$B(k,*)$}
      \State $C(*,*) \gets \alpha A(*,k) \cdot B(k,*) + C(*,*)$
    \EndFor
  \end{algorithmic}
  \label{alg:basesumma}
\end{algorithm}

DIMMA, an early variation of SUMMA introduced by Choi in 1998, improved performance of synchronous SUMMA by reducing the slack in communication \cite{Choi:1997cp,Choi:1998vw}.
The key insight of this work was that the order of broadcasts in the original SUMMA implementation, coupled with the communication barriers, created a significant slack in communication.
Iterations were reordered in DIMMA such that each node broadcasts all of its data in succession as opposed to the round-robin approach in SUMMA. Similar improvements can be attained by overlapping communication and computation \cite{Agarwal:1994hj}.

SUMMA was recently extended to sparse MM (SpSUMMA) by Bulu\c{c} and Gilbert \cite{Buluc:2008vn,Buluc:2012bp}.
This algorithm is similar to dense SUMMA, except matrix sub-blocks are stored in a doubly compressed sparse column (DCSC) format and sparse generalized matrix-multiplication (SpGEMM) is used to compute rank-$k$ updates.
The main challenges of all sparse MM algorithms, including SpSUMMA, are the increased relative costs of communication compared to the dense case, load imbalance, and the relatively low intranode performance of sparse matrix kernels.

The problem of load imbalance does not appear in dense SUMMA implementations as the work load is nearly-optimally balanced.  With random sparsity approximate load balance is achieved in the asymptotic limit, however with structured sparsity (e.g. matrices with decay) one does not expect rows/columns to be uniformly filled.

In the regime of high sparsity (low matrix fill-in factors) the communication time dominates the computation time due to the effectively-reduced benefit of blocking.
The increasing role of communication in sparse MM can be somewhat alleviated by communication hiding. Bulu\c{c} and Gilbert discussed potential benefits of communication
hiding for sparse MM in Ref. \cite{Buluc:2008vn} but did not pursue this approach in their experiments due to lack of quality one-sided communication tools \cite{Buluc:2012bp}.
Another possibility is to minimize communication, e.g. by switching to 3D \cite{Ballard:2013dy}.
Although sparse MM 2D SUMMA is not strongly scalable, nevertheless good scalability of SpSUMMA in practice was demonstrated \cite{Buluc:2012bp}.


\section{Task-based 2D SUMMA}

Our work to improve SUMMA is motivated by the needs of computation on matrices/tensors with irregular
block structure in the context of many-body quantum physics. Some of the challenges of computing with such data are similar to the general challenges of sparse MM: increased communication/computation ratio and lack of load balance. The latter is compounded by the desire to use physics-based blocking of matrix dimensions. To address these challenges we decided to investigate a task-based formulation of SUMMA algorithm, to partially offset some communication costs and to alleviate the load imbalance.
Prior efforts to reformulate dense matrix multiplication using a tasks-based programming models
are known \cite{pdsec,baruch:2001}; the novelty of our effort is the focus on dense matrices/tensors as well as matrices/tensor with irregular structure, such as block-rank-sparsity, that cannot be handled straightforwardly using the standard dense-only approaches.
In this section we describe the design of our algorithm, by highlighting the differences with the procedural SUMMA implementations\cite{VanDeGeijn:1997uf}
We first analyze the data dependencies of discrete operations in the procedural SUMMA implementation.
We then discuss the task composition and dependencies of our modified implementation.
For simplicity, we only consider the 2D SUMMA implementation; our approach is immediately applicable to the 2.5D variant since it's based on 2D SUMMA. 

\subsection{Dependency Analysis of Standard SUMMA}
\label{sec:ProceduralSUMMADependencyAnalysis}

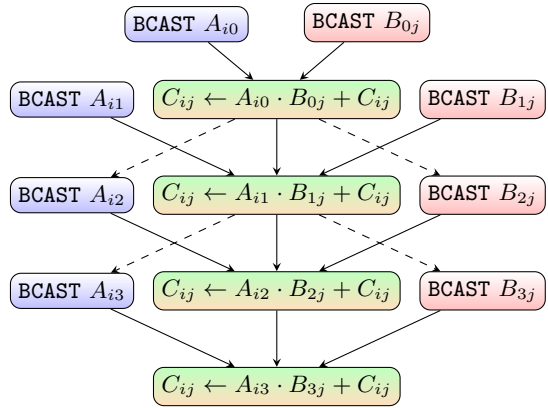
\begin{figure}[h!]
  \centering
  \begin{tikzpicture}[
  ->,
  >=stealth,
  node distance=0.25cm,
  task/.style={rectangle, draw, rounded corners=4pt,top color=white}
]
  
  \node[task,top color=green!25,bottom color=orange!25] (GEMM0) {$C_{ij} \gets A_{i0} \cdot B_{0j} + C_{ij}$};
  \node[task,bottom color=blue!25] (BCASTA0) [node distance=0.5cm,above=of GEMM0,xshift=-1.2cm] {{\tt BCAST} $A_{i0}$};
  \node[task,bottom color=red!25] (BCASTB0) [node distance=0.5cm,above=of GEMM0,xshift=1.2cm] {{\tt BCAST} $B_{0j}$};
  \node[task,top color=green!25,bottom color=orange!25] (GEMM1)   [node distance=0.75cm,below=of GEMM0] {$C_{ij} \gets A_{i1} \cdot B_{1j} + C_{ij}$};
  \node[task,bottom color=blue!25] (BCASTA1) [left=of GEMM0] {{\tt BCAST} $A_{i1}$};
  \node[task,bottom color=red!25] (BCASTB1) [right=of GEMM0] {{\tt BCAST} $B_{1j}$};
  \node[task,top color=green!25,bottom color=orange!25] (GEMM2)   [node distance=0.75cm,below=of GEMM1] {$C_{ij} \gets A_{i2} \cdot B_{2j} + C_{ij}$};
  \node[task,bottom color=blue!25] (BCASTA2) [left=of GEMM1] {{\tt BCAST} $A_{i2}$};
  \node[task,bottom color=red!25] (BCASTB2) [right=of GEMM1] {{\tt BCAST} $B_{2j}$};
  \node[task,top color=green!25,bottom color=orange!25] (GEMM3)   [node distance=0.75cm,below=of GEMM2] {$C_{ij} \gets A_{i3} \cdot B_{3j} + C_{ij}$};
  \node[task,bottom color=blue!25] (BCASTA3) [left=of GEMM2] {{\tt BCAST} $A_{i3}$};
  \node[task,bottom color=red!25] (BCASTB3) [right=of GEMM2] {{\tt BCAST} $B_{3j}$};

  \path (BCASTA0) edge node {} (GEMM0)
        (BCASTB0) edge node {} (GEMM0)
        (GEMM0)   edge node {} (GEMM1)
        (BCASTA1) edge node {} (GEMM1)
        (BCASTB1) edge node {} (GEMM1)
        (GEMM1)   edge node {} (GEMM2)
        (BCASTA2) edge node {} (GEMM2)
        (BCASTB2) edge node {} (GEMM2)
        (GEMM2)   edge node {} (GEMM3)
        (BCASTA3) edge node {} (GEMM3)
        (BCASTB3) edge node {} (GEMM3);

  \path [dashed]
           (GEMM0) edge node {} (BCASTA2)
           (GEMM0) edge node {} (BCASTB2)
           (GEMM1) edge node {} (BCASTA3)
           (GEMM1) edge node {} (BCASTB3);

\end{tikzpicture}
  \caption{A directed acyclic graph of the procedural SUMMA implementation consisting of broadcast ({\tt BCAST}) and rank-$k$ updates. Solid edges indicate data dependencies, and dashed edges indicate sequence dependencies.}
  \label{dag:naivesumma}
\end{figure}

Like all dense MM algorithms, SUMMA consists of data movement and computation tightly synchronized with each other (see Algorithm~\ref{alg:basesumma}).
Namely, the rank-$k$ update, $C_{ij} \gets A_{ik} B_{kj} + C_{ij}$, of each SUMMA iteration depends on the data from the broadcasts of the corresponding panels of {\bf A} and {\bf B} as well as the previous iteration's rank-$k$ update. In addition to these {\em data dependencies}, each broadcast is synchronized with a prior rank-$k$ update since communication operations are initiated at the beginning of each SUMMA iteration, as shown in Figure~\ref{alg:basesumma} (we denote such {\em sequence dependencies} by dashed lines).
Such design ensures that only a minimal memory is used (technically, nonblocking broadcasts require more memory than optimal).
However, such design limits the amount of parallelism available in a given iteration (see the  Figure~\ref{dag:naivesumma}).
Specifically, we can parallelize the rank-$k$ update of $C$ as well as the column and row broadcasts of $A$ and $B$, but SUMMA iterations --- although almost independent from one another --- are executed serially, one after the other. 
Furthermore, such design is not tolerant of any source of load imbalance, due to, for example, processor clock variation, network congestion, or --- most importantly for us --- due to the inhomogeneity of data.

\subsection {Task-Based Implementation}

To tolerate the data inhomogeneity, whether due to block sparsity or block-rank sparsity,
the task queue of each processes should have as many tasks as possible at any given time.
To maximize the amount of available parallelism, we introduced the following modifications
to the standard 2D SUMMA:
\begin{itemize}
\item data overdecomposition,
\item multiple-issue schedule of SUMMA iterations, and
\item task-decomposed broadcasts
\end{itemize}
Although the standard dense SUMMA allows near-perfect load balancing by
using element-cyclic embedding of matrices onto 2D mesh,
we over decompose matrices into arbitrarily sized blocks, also embedded cyclically.
In fact the dimension blocking is a central concept in {\sc TiledArray} library since it arises naturally in physical problems (hence the need to support arbitrary block sizes, including nonuniform blocking).
These blocks should be sufficiently small to ensure that each process is assigned {\em many} blocks,
yet large enough to perform acceptably with high performance BLAS3 library
(used to perform block-level multiplies).
Note that the overdecomposition also improves intra-node scalability
when coupled with an optimized, parallel runtime like Intel TBB,
and helps improve load balance in dealing with matrices/tensors that have permutational
symmetries, e.g. in the CAST algorithm \cite{Rajbhandari:2014bx}.

Decoupling of SUMMA iterations helps
to increase the amount of parallelism beyond what is afforded by a single SUMMA iteration.
This is similar to extracting more parallelism out of SUMMA by performing batches of SUMMA iterations in parallel on 2D subsets of a 3D process mesh in 2.5D MM algorithms \cite{Solomonik:2011bra}, except here different iterations of 2D SUMMA will be concurrently executed and scheduled on the same 2D mesh.
Note that the dependencies between SUMMA iterations occur via the result matrix {\bf C}.
To remove these data dependencies we assume that there is
enough memory available to split the rank-$k$ update operation into two separate tasks: a matrix-multiplication task producing a temporary block, $C_{ij}^{(k)} = A_{ik} \cdot B_{kj}$, and a reduction of the temporary into the result, $C_{ij} = C_{ij}^{(k)} + C_{ij}$.

\begin{figure}[h!]
  \centering
  \begin{tikzpicture}[
  ->,
  >=stealth,
  node distance=1.4cm,
  task/.style={rectangle, draw, rounded corners=4pt,top color=white}
]
  \foreach \x in {0,...,3}
  {
    \node[task,bottom color=green!25] at (0,-2cm*\x) (GEMM\x) {$C_{ij}^{(\x)} \gets A_{i \x} \cdot B_{\x j}$};
    \node[task,bottom color=blue!25] (BCASTA\x) [node distance=0.5cm,above=of GEMM\x,xshift=-1.2cm] {{\tt BCAST} $A_{i \x}$};
    \node[task,bottom color=red!25] (BCASTB\x) [node distance=0.5cm,above=of GEMM\x,xshift=1.2cm] {{\tt BCAST} $B_{\x j}$};

    \path 
       (BCASTA\x) edge node {} (GEMM\x)
       (BCASTB\x) edge node {} (GEMM\x);
  }

  \node[task,dashed,label=Parallel Reduce,minimum size=1cm,xscale=3.25, yscale=7.25] at (4cm,-3cm) (reduce) {};

  \foreach \x in {0,...,3}
  {
    \node[task,bottom color=orange!25] at (4cm,-2cm*\x) (SUM\x) {$C_{ij} \gets C_{ij}^{(\x)} + C_{ij}$};

    \path (GEMM\x) edge node {} (SUM\x);
  }


\end{tikzpicture}
  \caption{A direct translation of SUMMA to task-based implementation.}
  \label{dag:parallelsumma}
\end{figure}
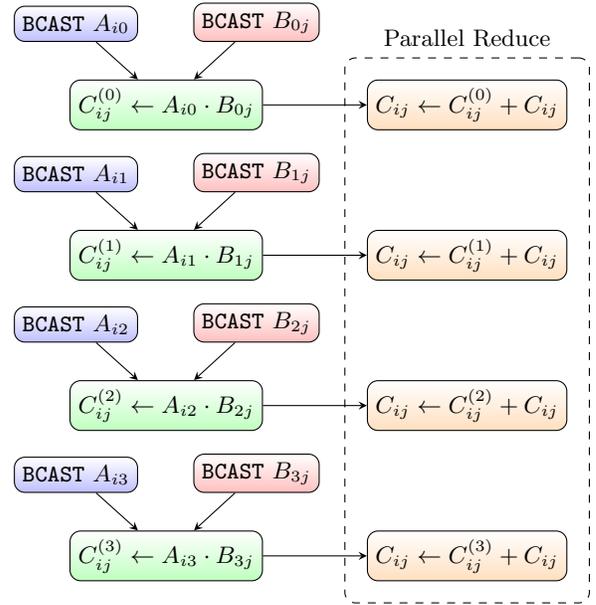

Figure~\ref{dag:parallelsumma} shows the directed acyclic graph (DAG) of task-based SUMMA with inter-iteration dependencies removed. 
By scheduling multiple iterations at once, we are able to overlap and interleave communication and computation of different iterations.
Also note that, in principle, all iterations of SUMMA can potentially be scheduled concurrently.
However, in practice we limit the number of {\em multiple-issue scheduling} to several iterations of SUMMA, and schedule additional iterations as the preceding iterations are retired, in a pipelined fashion.
The additional task dependencies needed to implement this {\em throttling} mechanism are not shown in Figure~\ref{dag:parallelsumma}.
The number of concurrent iterations is determined by the dimensions of the process grid.
For example, given a process grid with $P_{\rm row}$ rows and $P_{\rm col}$ columns, the number of concurrent iterations, $I$, is equal to,
\begin{equation}
 I(P_{\rm row},P_{\rm col};K) =
  \begin{cases}
   2             \text{  , if } P_{\rm row} < 2 \text{ or } P_{\rm col} < 2 \\
   K             \text{  , if } P_{\rm row} \ge K \text{ and } P_{\rm col} \ge K \\
   \min(P_{\rm row},P_{\rm col}) 
  \end{cases}
  \label{eqn:SummaItLimit}
\end{equation}
where $K$ is the number of blocks along the inner dimension of the matrix multiplication.
This limit is based on performance profiling, where we observed the maximum performance at $\min(P_{\rm row},P_{\rm col})$.

Lastly, we decomposed broadcast tasks into subtasks each of which deals with a point-to-point
communication. This achieves benefits similar to what pipelining does to standard 2D SUMMA \cite{VanDeGeijn:1997uf,Tan:2013fl},
by introducing the overlap between communication and computation of tasks working
on a given SUMMA iteration. For example, a given process can start doing compute work as soon as the minimal amount of data --- one block of {\bf A} and one block of {\bf B} --- is available, concurrently with any additional communication of these blocks to other processes in the row and column groups.
Work scheduling based on the availability of data in our task approach should improve tolerance to latency and network topology \cite{Nguyen:2012vu}. In the current implementation we use a hardware-oblivious binary tree broadcast, {\em i.e.}\ it is not optimized for any particular interconnect hardware or topology, such as a multidimensional torus or fat tree.
While our design does not preclude the use of optimized broadcast algorithms, a hardware oblivious algorithm should help performance portability across high-end and commodity
hardware.

\begin{figure}[h!]
  \centering
  \begin{tikzpicture}[
  ->,
  >=stealth,
  node distance=1.4cm,
  task/.style={rectangle, draw, rounded corners=4pt,top color=white}
]

  \def\rows{1}   
  \def\cols{1}   
  \def\rowsize{3.3cm}
  \def\colsize{-2.15cm}
  
  \foreach \y in {0,...,\cols}
    \node[task,bottom color=blue!25] at (0cm,\colsize*\y) (BCASTA\y) [yshift=-1cm] {{\tt BCAST} $A_{i k}^{(\y)}$};

  \foreach \x in {0,...,\rows}
    \node[task,bottom color=red!25] at (\rowsize*\x,0cm) (BCASTB\x) [xshift=1.15cm] {{\tt BCAST} $B_{k j}^{(\x)}$};

  \foreach \x in {0,...,\rows}
    \foreach \y in {0,...,\cols}
  {
      \node[task,bottom color=green!25] at (\rowsize*\x,\colsize*\y) 
          (GEMM-\x-\y) [xshift=2.8cm,yshift=-1.5cm] {$C_{ij}^{(\y,\x)} \gets A_{i k}^{(\y)} \cdot B_{k j}^{(\x)}$};
      \node[task,bottom color=orange!25]         
          (SUM-\x-\y) [node distance=0.25cm,below=of GEMM-\x-\y] {$C_{ij} \gets C_{ij}^{(\y,\x)} + C_{ij}$};

      \draw (BCASTA\y)    -| (GEMM-\x-\y);
      \draw (GEMM-\x-\y)  -- (SUM-\x-\y) ;
  }

  \foreach \x in {0,...,\rows}
    \foreach \y in {0,...,\cols}
      \draw[line width=3pt,white] (BCASTB\x) |- (GEMM-\x-\y);

  \foreach \x in {0,...,\rows}
    \foreach \y in {0,...,\cols}
      \draw (BCASTB\x) |- (GEMM-\x-\y);

\end{tikzpicture}
  \caption{A directed acyclic graph for a single, decomposed SUMMA iteration. Broadcast, matrix multiply, and reduction tasks are decomposed such that each sub-block of the local result matrix, $C_{ij}^{(x,y)}$, depends on two sub-block broadcast tasks for $A_{ik}^{(x)}$ and $B_{kj}^{(y)}$.}
  \label{dag:summaiter}
\end{figure}
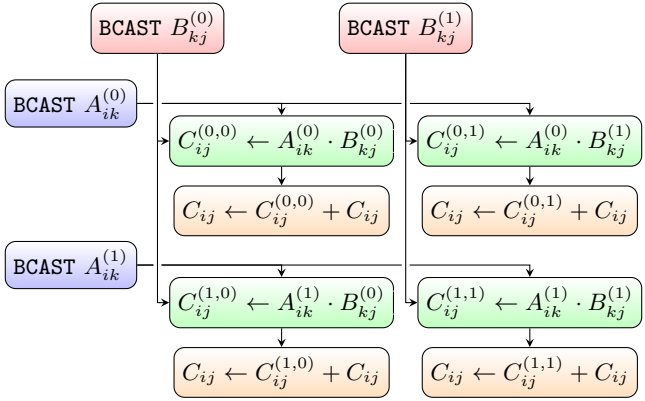

The dynamic scheduling of computation and communication trades off the predictable bounds on the resource use, in particular memory and network bandwidth, for high performance. For example, by multiple-issue schedule of SUMMA iterations increases memory use because each node must store an additional column/row block of the the argument matrices. Additional temporary storage might be used by matrix-multiplication producing temporaries subsequently reduced into the blocks of the target.
The average memory overhead per node per SUMMA iteration, as depicted in Figure~\ref{dag:parallelsumma}, is proportional to:
\begin{equation}
  \underbrace{ \frac{Mmk}{P_{\rm row}}}_{\text{ left-hand }} +
  \underbrace{\frac{Nkn}{P_{\rm col}}}_{\text{ right-hand }} +
  \underbrace{\frac{MNmn}{P_{\rm row} P_{\rm col}}}_{\text{ result }}
\end{equation}
where $k$ is the average block size of the inner dimension; $m$ and $n$ are the average number of rows and columns in the result matrix blocks, respectively; $M$ and $N$ are the number of block rows and block columns in the result matrix, respectively; and $P_{\rm row}$ and $P_{\rm col}$ are the number of rows and columns in the process grid, respectively.
However, we mitigate the memory overhead associated with the local result matrix by decomposing the rank-$k$-update tasks such that each matrix-multiplication task only computes a small sub-block of the local result matrix (see Figure~\ref{dag:summaiter}).
This decomposition of the rank-$k$-update and broadcast operations are similar to the decomposition of work between nodes within a SUMMA iteration, but without the analogous communication (dependencies) between computation tasks.
This allows different threads to work on different sub-blocks of the local result matrix concurrently.
The maximum memory overhead is therefore bounded by the number of threads and only weakly dependent on the number of concurrent iterations.
We further mitigate memory costs by combining the matrix-multiplication and reduction tasks whenever possible, which  eliminates the memory and computational overhead of a separate reduction task.
As a result, matrix-multiplication tasks only allocate additional, temporary memory for the result sub-blocks when two or more threads compute contributions to the same sub-block simultaneously.
The memory cost associated with the left- and right-hand matrices is also mitigated by the overdecomposition of data.
This is due to the fact that each row and column sub-block only remains in memory as long as it is needed, which is not necessarily for the entire duration of computation for an iteration.
Finally, the overall memory consumption can be controlled by limiting the number of concurrent SUMMA iterations that are executed on each node. If $I$ SUMMA iterations are scheduled at once each process's memory potentially increases over that of standard SUMMA by a factor of $I$; although as explained the resulting increase would in practice be much less than that.

\section{Results}

The task-based 2D SUMMA algorithm is implemented in C++ in the {\sc TiledArray} library \cite{TiledArray}.
{\sc TiledArray} uses the {\sc MADNESS} runtime \cite{MADNESS} to express the computation in terms of tasks and to manage the low-level details of task scheduling and data movement. Both {\sc TiledArray} and {\sc MADNESS} can be obtained under the terms of the GNU General Public License.

\subsection{Experimental Setup}

In our experiments, we compare the performance of the task-based SUMMA implementation for matrix multiplication of square uniformly- and nonuniformly-blocked real double-precision matrices;
the square shape was chosen for simplicity of the performance analysis and is not a constraint of our implementation.
The Basic Linear Algebra Subprograms (BLAS) DGEMM routine was used as the block multiply-add kernel in our MM tasks.
On the BG/Q system, we use the BLAS functions in IBM's Engineering and Scientific Subroutine Library (ESSL).
Intel's Math Kernel Library (MKL) is used on the small, x86\_64 Linux cluster.
Although concurrent versions of DGEMM are available in both libraries, we use serial DGEMM to properly allocate resources with the MADNESS task-based parallel runtime system; this limitation will be removed in the near future. Reported wall times are an average of $n$ repeated multiplications of two random matrices, where $n=30$ for most computations except when run time limits constrained the sample sizes.

Block size for the uniform MM tests are optimized for each system, where we select the block size with the best performance from a series of single-process MM tests with a range of block sizes.
To determine the size of blocks in the nonuniform MM tests, we first start by constructing $M$ empty, row blocks, where $M$ is equal to the number of row blocks in the uniformly-blocked matrices.
We then randomly one to the size a row block, and repeat this step until the total number of rows among all blocks is equal to the number of rows in the uniformly blocked matrices.
We repeat procedure for column block sizes.
This ensures the average block size, as well as the number of blocks per matrix, in the nonuniform MM tests are equal to that of the uniformly-blocked matrices.

\subsection{Performance of Task-Based SUMMA}
\begin{figure}[h!]
  \centering
  \includegraphics[width=\columnwidth]{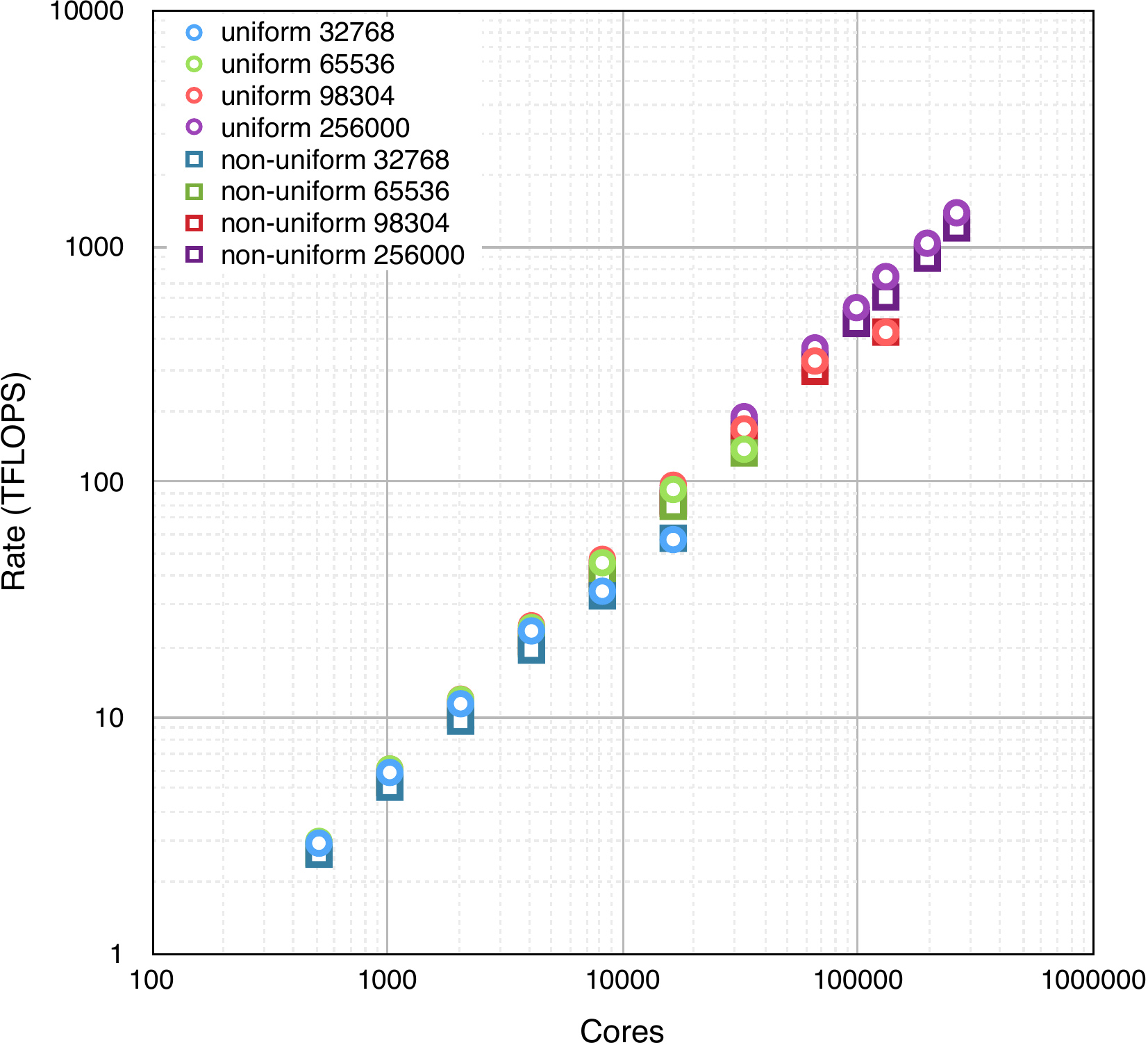}
  \caption{Floating-point operation rate for multiplication of uniformly- and nonuniformly-blocked square matrices on IBM BG/Q. The matrix sizes included in the test are $32,768$, $65,536$, $98,304$, $256,000$.}
    \label{fig:bgqspeed}
\end{figure}
\begin{figure}[h!]
  \includegraphics[width=\columnwidth]{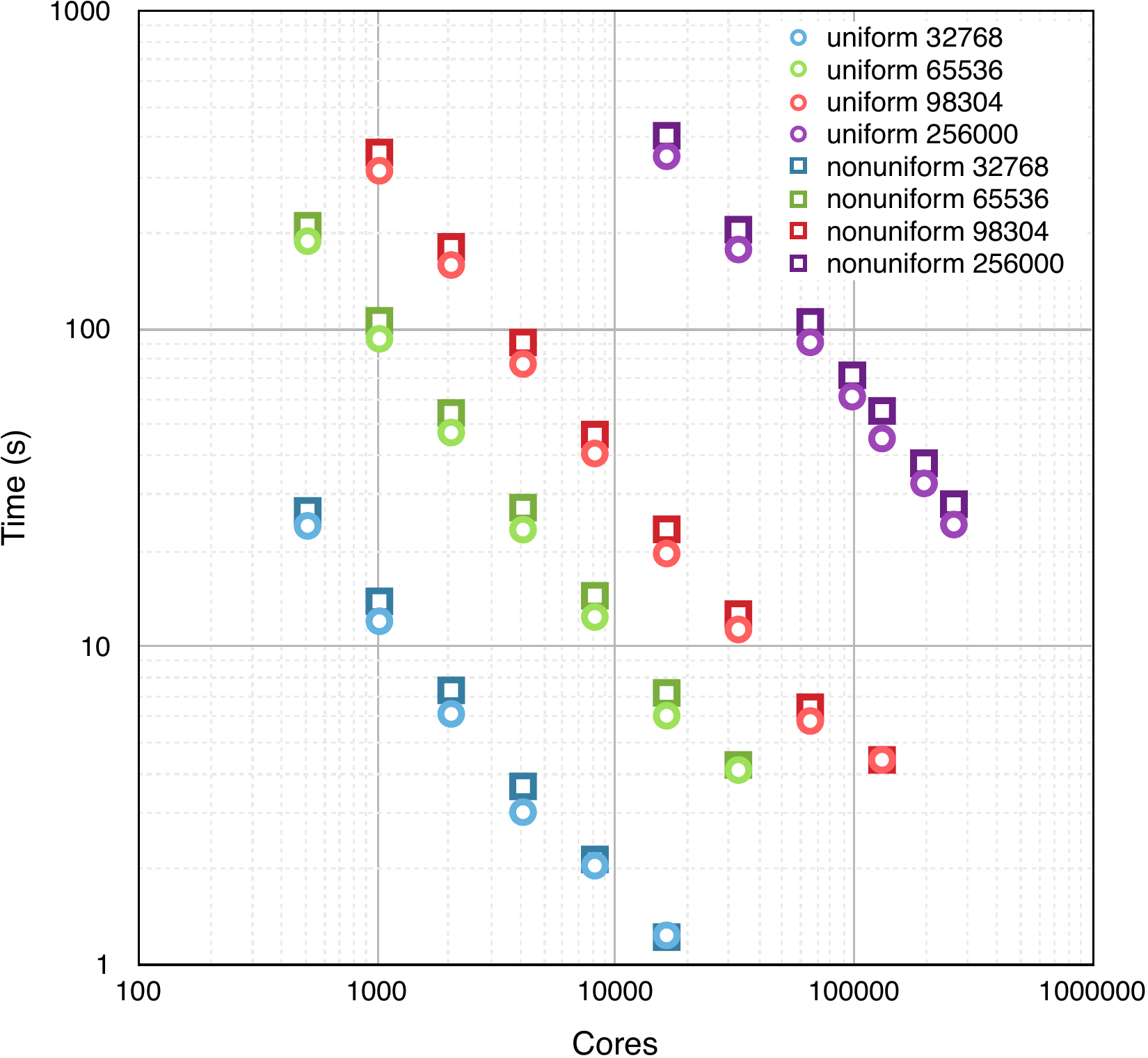}
  \caption{Wall time for multiplication of uniformly- and nonuniformly-blocked square matrices on IBM BG/Q. The matrix sizes included in the test are $32,768$, $65,536$, $98,304$, $256,000$.}
    \label{fig:bgqtime}
\end{figure}

Performance measurements on high-end hardware were performed on ``Mira,'' an IBM Blue Gene/Q  (BG/Q) system at Argonne National Laboratory.
For each matrix size we varied the number of compute nodes from $2^{5}$ and $2^{14}$ (each node has 16 compute cores with 4 hardware threads per core; thus the largest computation utilized $2^{20} \approx 10^{6}$ threads).

In Figure~\ref{fig:bgqspeed} and~\ref{fig:bgqtime}, we see an approximately linear increase in computational rate for each matrix size.
The combine data for all matrix sizes also shows approximately linear weak scaling in the entire range of processor counts.
The performance difference between  the uniform and nonuniform tests is small and follow each other closely, though the tests with uniform block sizes are consistently faster than those with nonuniform block sizes.
In fact, slopes for both the uniform and nonuniform rate tests are approximately equal.
This suggests that the performance difference between these two tests is due to an intra-node effect, rather than a scaling issue that one would expect to see with a significant load imbalance.
This difference is an expected result since DGEMM performance is sensitive to the matrix size for small matrices.

\begin{figure}[h!]
  \centering
  \includegraphics[width=\columnwidth]{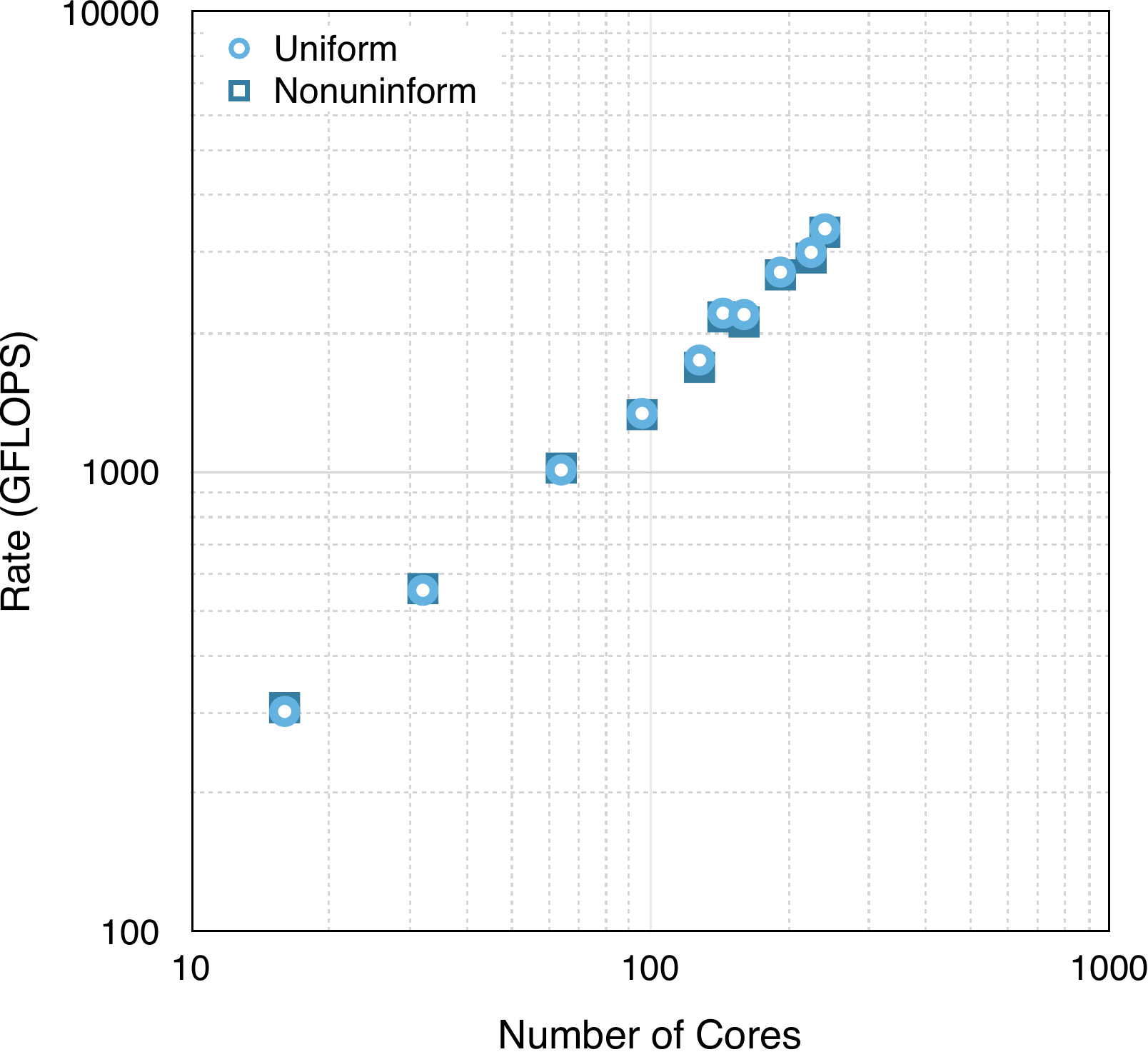}
  \caption{Floating operation throughput for multiplication of uniformly- and nonuniformly-blocked square matrices of size $32,768$ on a commodity cluster.}
  \label{fig:hyadesspeed}
\end{figure}
\begin{figure}[h!]
  \includegraphics[width=\columnwidth]{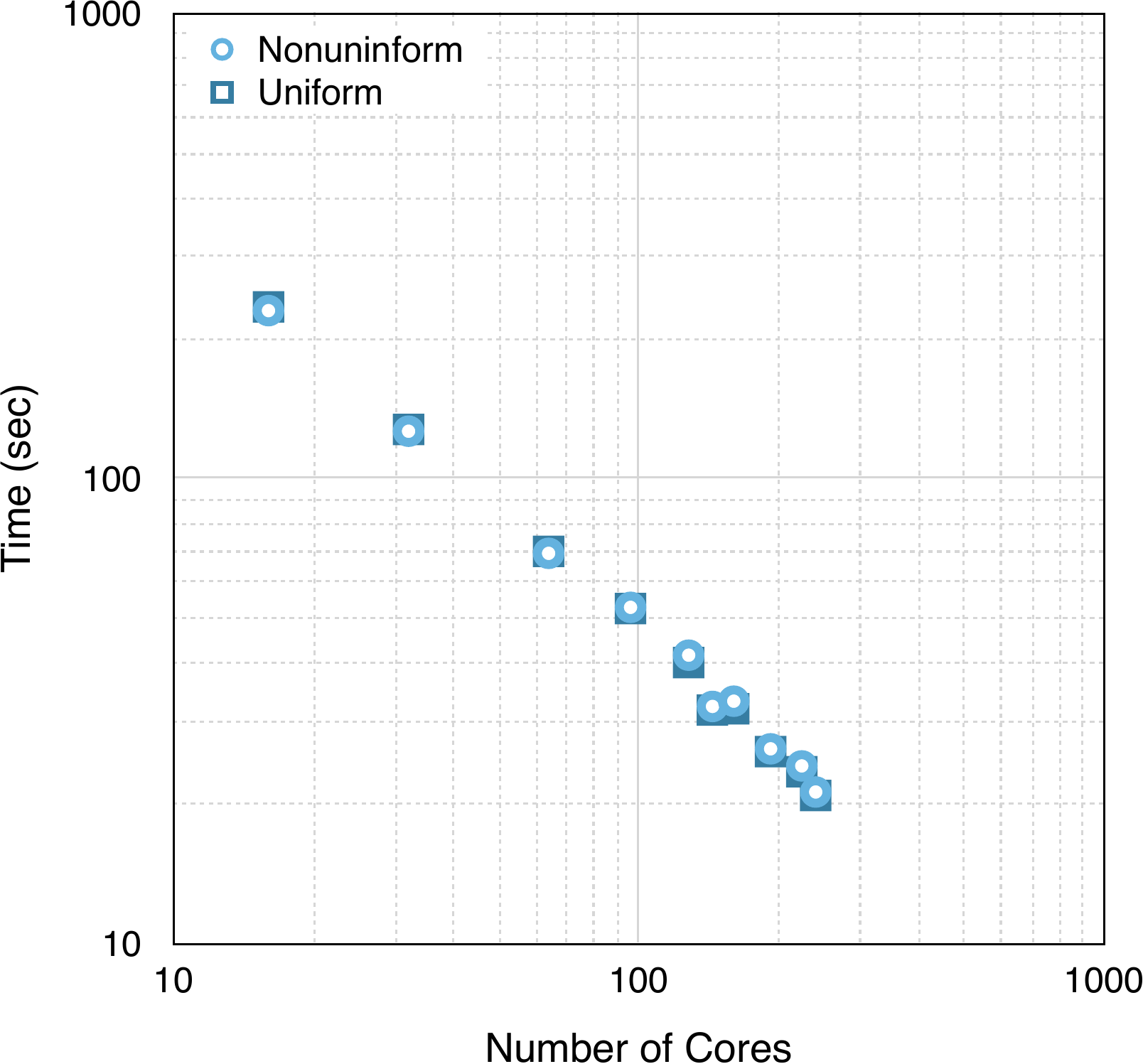}
  \caption{Wall time for multiplication of uniformly- and nonuniformly-blocked square matrices of size $32,768$ on a commodity cluster.}
  \label{fig:hyadestime}
\end{figure}
We also perform a strong scaling test on a small, fifteen-node cluster with two eight-core 2.60 GHz Intel Xeon E5-2670 processors (16 cores total) and 16 GB of RAM per node, with an Infiniband switch.
We use a matrix size of $32,768$ for all tests on this system.
The block size is set to $256$.
Similarly, the average number of rows and columns per block in the nonuniform test was 256 for each dimension. The number of compute cores in this test set varies from 16 to 240.

Like the tests on the BG/Q system, the data shows an approximately linear increase in computational rate for both uniform and nonuniform tests (see Figures~\ref{fig:hyadesspeed} and~\ref{fig:hyadestime}).
Unlike the previous tests, the performance difference between the uniform and nonuniform tests is negligibly small; neither set of tests clearly out performs the other.
For example, the mean computational times of the fifteen-node, uniform- and nonuniform-matrix tests are 21.1716 and 20.8096 seconds, respectively, with standard deviations of 0.663725 and 0.544411 seconds, respectively.
The mean values of these data points are within one standard deviation of each other, and other data points with the same number of processes are similarly close.

\subsection{Efficiency of Task-Based SUMMA}

The performance and efficiency of our task-based SUMMA implementation was found to be highly dependent on that of the matrix multiplication kernel, in this case DGEMM ({\sc TiledArray} supports matrices of standard single- and double-precision real and complex types as well as user-defined types).
The best, single-node performance we achieved with ESSL is 99.1393 GFLOPS, which is 59.7\% of Rmax (166.06 GFLOPS per-node) and 48.4\% of Rpeak (204.799 GFLOPS per-node).
Typically, with a well optimized BLAS library, single-node performance of our task-based SUMMA implementation can achieve 90\% to 95\% of theoretical peak.
For example, the single node performance on our small cluster, using MKL, was 302.787 GFLOPS, which is 90.98\% of the theoretical peak performance computed from the base CPU frequency (332.8 GFLOPS per node).
The discrepancy in percent of machine peak between these two systems is due to the limited thread scalability of ESSL. Anecdotally, 50\% of machine peak is considered ``good'' performance for ESSL BLAS Level-3 functions with 16 cores.

\begin{figure}[h!]
  \centering
  \includegraphics[width=\columnwidth]{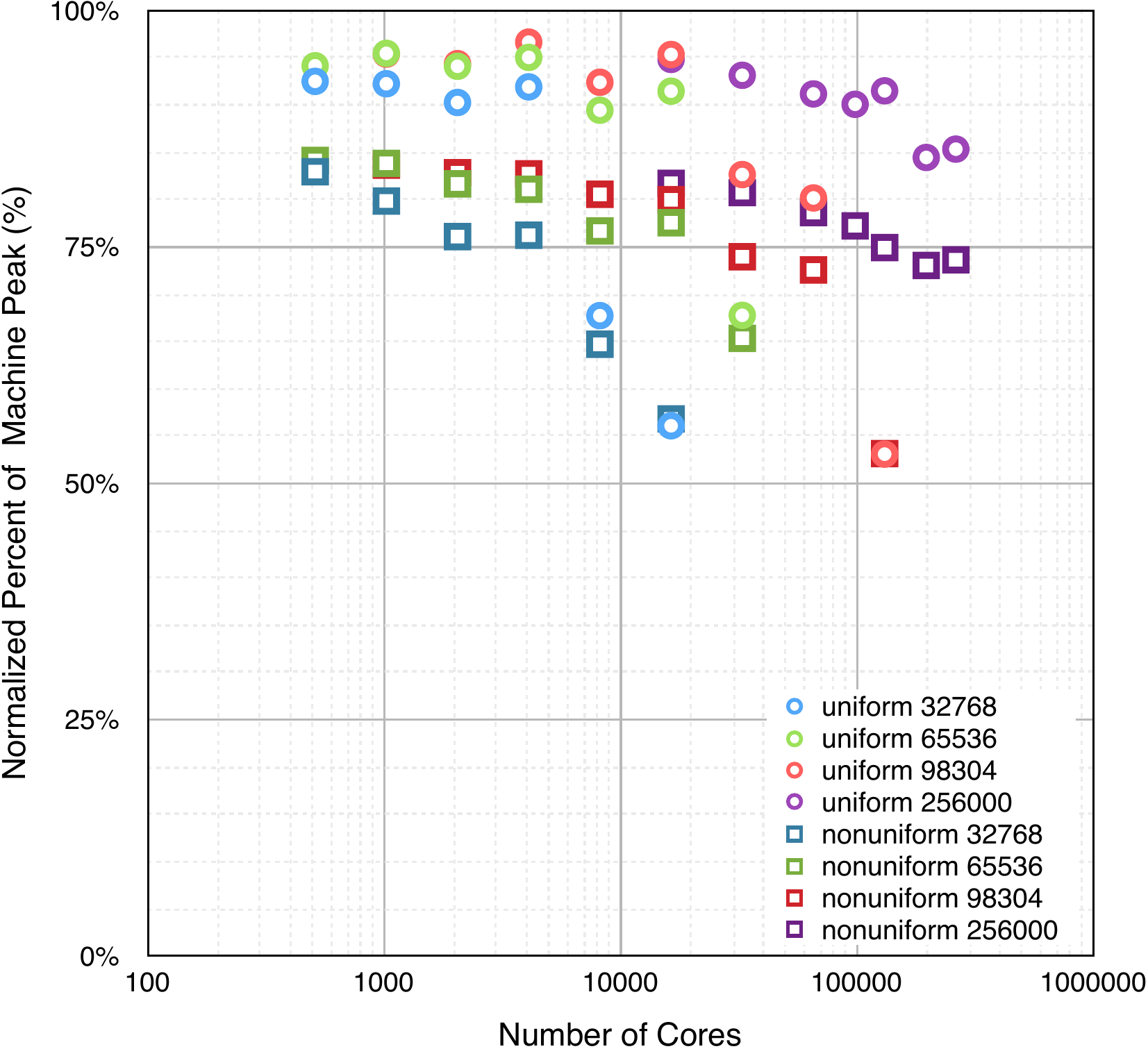}
  \caption{Computational efficiency for multiplication of uniformly- and nonuniformly-blocked square matrices on IBM BG/Q. One hundred percent efficiency refers to the peak, measured performance of a single node, with {\sc TiledArray}.}
  \label{fig:bgqefficiency}
\end{figure}

In Figure~\ref{fig:bgqefficiency}, we show the percent of efficiency of our algorithm on the IBM BG/Q system, with 100\% efficiency set at the maximum measured performance on a single node (99.1393 GFLOPS).
The data shows that usually greater than 75\% efficiency is maintained across a variety of matrix sizes and number of cores, relative to the single node performance. The efficiency drops down to as low as 50\% in the strong scaling limit, when each processor core has 4 or fewer target matrix blocks. 
The granularity of tasks limits the maximum amount of parallelism achievable for each matrix size, which causes a steep drop in efficiency in the right-most tail of each data set.

\subsection{Load Variability with Nonuniform Blocking}

The randomly sized matrix blocks in these tests are intended to simulate the irregular blocking structure of data in quantum physics problems.
We used a low-quality random number generator to create significant inhomogeneity in the blocks sizes.
The ratio of minimum to maximum memory and work loads for these generated matrices are given in Table~\ref{tab:nonuniformminmax}.
\begin{table}[h!]
  \centering
  \begin{tabular}{ |p{0.2\columnwidth}|p{0.2\columnwidth}|p{0.2\columnwidth}| }\hline
    Matrix Size & Memory Min:Max & Work Min:Max \\ \hline
    32768         & 1:2.99                  & 1:4.46 \\ \hline
    65536         & 1:3.54                  & 1:5.77 \\ \hline
    98304         & 1:3.27                  & 1:5.51 \\ \hline
    256000       & 1:3.93                  & 1:7.12 \\ \hline
  \end{tabular}
  \caption{Ratio of the minimum to maximum memory and work loads for nonuniformly-blocked matrices.}
  \label{tab:nonuniformminmax}
\end{table}

Despite the substantial inhomogeneity at the block level, because each process hold many blocks due to data overdecomposition, the effective load imbalance in the computation and communication is somewhat smaller than what is reported in Table~\ref{tab:nonuniformminmax}.
For example, with a matrix size of $32,768$ and 256 processes, we see a ratio of 1:1.35 between processes with the smallest and largest memory usage.

\section{Conclusions}
We presented a fine-grained task-based reformulation of Scalable Universal Matrix Multiplication Algorithm (SUMMA). In conjunction with multiple-issue scheduling of SUMMA iterations, the new formulation
should be tolerant of data inhomogeneity due to matrix structure.
The implementation of our algorithm in {\sc TiledArray} library performed equally well for  multiplication of square matrices with uniform and nonuniform blocking (the latter designed to simulate the domain-specific blocking structure characteristic of the electronic structure domain).
The implementation scales well on a small commodity cluster as well as a high-end IBM BG/Q supercomputer, with typical measured efficiencies relative to the single node performance of 75\% on as many as $2^{18}$ cores. 

\section{Acknowledgments}
This work was supported by the National Science Foundation (grants CHE-1362655 and ACI-1047696), Camille Dreyfus Teacher-Scholar Award, and Alfred P. Sloan Fellowship.
We gratefully acknowledge the computer time provided by the Innovative and Novel Computational Impact on Theory and Experiment (INCITE) program. This research used resources of the Argonne Leadership Computing Facility, which is a DOE Office of Science User Facility supported under Contract DE-AC02-06CH11357. The use of commodity cluster was acquired with the help of the U.S. National Science Foundation award CHE- 0847295. We also acknowledge the started allocation from Advanced Research Computing at Virginia Tech.

We would like to thank Mr. Cannada A. Lewis for the help with data analysis and useful contributions to {\sc TiledArray}.

%
\bibliographystyle{hacm}
\bibliography{refs,refs-ev}  
%
%

\end{document}